\documentclass[3p]{elsarticle}

\usepackage[colorlinks, hyperindex]{hyperref}
	\hypersetup
	{
		colorlinks, %
		citecolor=blue, %
		linkcolor=Green, %
		urlcolor=blue, %
	}

\usepackage{amsmath}
           \usepackage{mathrsfs}
	\usepackage{amssymb}
	\usepackage{graphicx} 
	\usepackage{makeidx}
	\usepackage{amsfonts}
	\usepackage[ansinew]{inputenc}
	\usepackage[usenames, dvipsnames]{pstricks}
	\usepackage{subfigure}
	\usepackage{epsfig}

\journal{arXiv}

\begin{document}

\begin{frontmatter}

\title{Physical Aspects of Unitary evolution of Bianchi-I Quantum Cosmological Model}

\author{Sridip Pal\fnref{myfootnote}}
\ead{sridippaliiser@gmail.com}
\address{Department of Physics,
University of California, San Diego,\\
9500 Gilman Drive, La Jolla\\ CA 92093, USA}
\fntext[myfootnote]{Work partially done in: Department of Physical Sciences, 
Indian Institute of Science Education and Research Kolkata, 
Mohanpur, West Bengal 741246, India}

\begin{abstract}
In this work, we study some physical aspects of unitary evolution of Bianchi-I model. In particular, we study the behavior of the volume and the scale factor as a function of time for the Bianchi-I universe with ultra-relativistic fluid ($\alpha=1$). The expectation value of volume is shown not to hit any singularity. We elucidate on the anisotropic nature of the solution and physically interpret the wavefunction as a superposition of collapsing universe and expanding universe mimicking Hartle-Hawking type wavefunction. The same analysis has been done for $\alpha\neq 1$ as well. The work also  comments on how to obtain Vilenkin type wavefunction representing an expanding universe without that being superposed with a solution representing a collapsing universe. We also point at a possible phase transition phenomenon for $\alpha\neq1$ case with existence of a critical parameter.
\end{abstract}


\end{frontmatter}

\section{Introduction}
Quantum Cosmology, an amalgamation of Quantum theory with Cosmology, deals with quantizing a cosmological model as a whole. The quantization scheme builds upon the famous Wheeler-DeWitt Equation \cite{dewitt, wheeler}  paving the way to have a hold on the physics of the beginning of the universe. However, we have several conceptual and technical problems associated with the formulation. One of the important problems involves suitably choosing a time like parameter since in a relativistic theory, time itself is a coordinate \cite{kuchar1, isham, anderson}. In another study \cite{rovelli}, a scheme has been proposed where time plays no role whatsoever. Apart from this,  there are conceptual problems regarding the interpretation of the wave function, implementation of proper boundary conditions,  and several others. The various aspect of these subtleties have been reviewed along with the development of the subject in \cite{wilt, halli, nelson1}.\\

The issue of conjuring up a time parameter can be resolved if matter sector comprises a perfect fluid, a scenario, which enables us to exploit a formalism, relating the four velocity of the fluid with thermodynamic parameters, developed by Schutz \cite{schutz1, schutz2}. It so happens that after some clever canonical transformations one can show that the canonical momentum associated with fluid variable appears as a linear term in Hamiltonian and thereby mimics Schrodinger equation. Hence the variable canonically conjugate to this momentum can be treated as the time parameter. Lapchinskii and Rubakov\cite{rubakov} utilized this strategy to identify a well behaved time parameter. This method has found a renewed and very successful application by Alvarenga and Lemos\cite{alvarenga1} and has subsequently been used, for the quantization of an isotropic cosmological model, in the works of Batista et al\cite{batista}, Alvarenga et al\cite{alvarenga2} and Vakili\cite{vakili1, vakili2}. The method also finds an application in quantizing anisotropic models in the work done by Alvarenga et al\cite{alvarenga3}, Majumder and Banerjee\cite{barun}, Pal and Banerjee\cite{sridip}.\\

One of the pathology that came up while employing the method to anisotropic universe is the non unitarity associated with evolution operator, which is a result of having a non self-adjoint Hamiltonian. This pathology was present in the Bianchi I model investigated by Alvarenga et. al\cite{alvarenga3}. It should be mentioned that without matter sector, we do not have any physical identification of time, hence the non unitarity associated with the model might escape notice \cite{lidsey, nelson2} while the use of fluid evolution enables us to study the issue critically.\\

Majumder and Banerjee \cite{barun} showed that a suitable operator ordering can make the norm of the wave function, corresponding to an anisotropic model, time independent in the asymptotic limit. Although as a whole the model still suffered from non unitarity, it was a good enough indication that the problem of non unitarity has to do with operator ordering rather than being a generic feature. The confirmation comes in a later investigation\cite{sridip}, where it has been shown, contrary to the folklore, that the non unitarity is indeed not a generic feature of anisotropic cosmologies. With a proper choice of operator ordering, unitarity can be restored to the model. Initially it was shown for Bianchi-I model, later the work has been extended to encompass the Bianchi-V and Bianchi-IX universe as well\cite{sridip2}.\\

This work sheds light on some physical aspects of ``Unitarity Restored Bianchi-I Model" building upon the work in \cite{sridip,sridip2}. We elucidate on the nature of wavefunction along with a critical study on evolution of scale factors, volume, manifest anisotropy of the solutions to the model. The physical insight strengthens the idea presented earlier in \cite{sridip,sridip2} and leads to a better understanding of the physics of a quantized anisotropic model.\\

The paper is organized as follows. In section two, Schutz formalism is briefly described along with quantization of Bianchi-I model and a discussion on the evolution of expectation value of volume, the scale factors associated with the model. In Section three, we take up the $\alpha=1$ case in detail and show that it is possible to have a quantized unitary Bianchi-I model with an explicit anisotropy. In section four, we discuss the $\alpha\neq1$ case, elucidate on the nature of the wavefunction and the eigenspectra of energy. We also point at a possible phase transition with existence of a critical parameter. The fifth and final section discusses the results obtained in the work and its implications.

\section{Quantization of Bianchi-I Cosmology}
The general action for gravity is given by
\begin{equation}\label{Action}
\mathcal{A}=\int_{M}d^{4}x \sqrt{-g}R +2 \int_{\partial M} \sqrt{h}h_{ab}K^{ab}+\int_{M} d^{4}x \sqrt{-g}P,
\end{equation}
where $K^{ab}$ is the extrinsic curvature, and $h^{ab}$ is the induced metric over the boundary $\partial M$  of the 4 dimensional space-time manifold $M$. The units are so chosen that $16\pi G = 1.$ 

The metric for Bianchi-I is given by
\begin{equation}\label{Metric}
ds^{2}=n^{2}dt^{2}-\left[a^{2}(t)dx^{2}+b^{2}(t)dy^{2}+c^{2}(t)dz^{2}\right],
\end{equation}
where $n(t)$ is the lapse function and $a,b,c$ are functions of the cosmic time $t$. 
With this metric the gravity sector of \eqref{Action} takes the form
\begin{equation}\label{GRaction}
\mathcal{A}_g =\int dt \left[-\frac{2}{n}\left(\dot{a}\dot{b}c+\dot{b}\dot{c}a+\dot{c}\dot{a}b\right)\right].
\end{equation}
With  a set of new variables
\begin{eqnarray}\label{coordinate}
a(t)&=&e^{\beta_{0}+\beta_{+}+\sqrt{3}\beta_{-}},\\
b(t)&=&e^{\beta_{0}+\beta_{+}-\sqrt{3}\beta_{-}},\\
c(t)&=&e^{\beta_{0}-2\beta_{+}},
\end{eqnarray}
the Lagrangian density of the gravity sector becomes
\begin{equation}\label{7}
\mathcal{L}_{g}=-6\frac{e^{3\beta_{0}}}{n}\left(\dot{\beta}_{0}^{2}-\dot{\beta}_{-}^{2}-\dot{\beta}_{+}^{2}\right).
\end{equation}
The corresponding Hamiltonian density can be written as
\begin{equation}\label{hamgrav}
 H_{g}=-n\exp(-3\beta_{0})\left\{\frac{1}{24}\left(p_{0}^{2}-p_{+}^{2}-p_{-}^{2}\right)\right\}.
\end{equation}
Now the  Schutz's formalism \cite{schutz1, schutz2} is employed to identify a time parameter out of the matter sector, which comprises an ideal fluid given by an equation of state $P = \alpha \rho$ where $\alpha$ is a constant (with $\alpha \leq 1$), $P$ and $\rho$ is the pressure and density of the fluid respectively. Standard thermodynamical consideration leads to the following action for the matter sector,
\begin{equation}
\label{3.91}
\begin{split}
\mathcal{A}_{f}&=\int dt \mathcal{L}_{f}\\&= V\int dt \left[n^{-\frac{1}{\alpha}}e^{3\beta_{0}}\frac{\alpha}{\left(1+\alpha\right)^{1+\frac{1}{\alpha}}}\left(\dot{\epsilon}+\theta\dot{S}\right)^{1+\frac{1}{\alpha}}e^{-\frac{S}{\alpha}}\right],
\end{split}
\end{equation}
where $\theta$, $\epsilon$ and S is defined as in the reference \cite{sridip, sridip2}.	
We define canonical momentum to be $p_{\epsilon}=\frac{\partial\mathcal{L}_{f}}{\partial\dot{\epsilon}}$ and
$p_{S}=\frac{\partial\mathcal{L}_{f}}{\partial\dot{S}}$ and the corresponding Hamiltonian density comes out to be
\begin{equation}
\label{hamfl}
H_{f}=ne^{-3\alpha\beta_{0}}p_{\epsilon}^{\alpha +1}e^{S}.
\end{equation}
With the following canonical transformation,
\begin{eqnarray}\label{9}
T&=&-p_{S}\exp(-S)p_{\epsilon}^{-\alpha -1},\\
p_{T}&=&p_{\epsilon}^{\alpha+1}\exp(S),\\
\epsilon^{\prime}&=&\epsilon+\left(\alpha+1\right)\frac{p_{S}}{p_{\epsilon}},\\
p_{\epsilon}^{\prime}&=&p_{\epsilon},
\end{eqnarray}
which maintains the canonical structure by preserving the Poisson commutation relation\cite{sridip}, the Hamiltonian density for the fluid sector can be cast into following form 
\begin{equation}
H_{f}= ne^{-3\beta_{0}}e^{3\left(1-\alpha\right)\beta_{0}}p_{T}.
\end{equation}

Combining gravity and fluid sector, the super Hamiltonian (to be precise, super Hamiltonian density) becomes
\begin{equation}
H=-\frac{ne^{-3\beta_{0}}}{24}\left(p_{0}^{2}-p_{+}^{2}-p_{-}^{2}-24e^{3\left(1-\alpha\right)\beta_{0}}p_{T}\right)
\end{equation}
and variation with respect to $n$ yields Hamiltonian constraint
\begin{equation}
\label{master}
 \mathcal{H}=\frac{H}{n}=\frac{1}{n}\left(H_{g}+H_{f}\right)=0.
\end{equation}
We now promote the variation of super Hamiltonian with respect to $n$ i.e $\mathcal{H}$ to an operator and postulate commutator relation amongst the quantum operators as usual i.e we have:
\begin{equation}\label{12}
p_{j}\mapsto -\imath\hbar\partial_{\beta_{j}},
\end{equation}
for $i=0,+$, and 
\begin{equation}\label{13}
p_{T} \mapsto -\imath \hbar\partial_{T}.
\end{equation}
This mapping is equivalent to postulating the fundamental commutation relations:
\begin{equation}\label{13.1}
\left[\beta_{j},p_{k}\right]=\imath\hbar\delta_{jk}\mathbb{I}.
\end{equation}

In all subsequent discussion, we employ the choice of natural unit i.e $\hbar=c=1$ and rewrite the equation \eqref{master} in its operator form as $\mathcal{H}\psi=0$, which can be recast, using the operator ordering as prescribed in \cite{sridip}, in following form:

\begin{equation}\label{WDeq}
\left[e^{\frac{3}{2}\left(\alpha -1\right)\beta_{0}}\frac{\partial}{\partial\beta_{0}}e^{\frac{3}{2}\left(\alpha -1\right)\beta_{0}}\frac{\partial}{\partial\beta_{0}} -e^{3\left(\alpha -1\right)\beta_{0}}\frac{\partial^{2}}{\partial\beta_{+}^{2}}-e^{3\left(\alpha -1\right)\beta_{0}}\frac{\partial^{2}}{\partial\beta_{-}^{2}}\right]\psi =24\imath\frac{\partial}{\partial T}\psi.
\end{equation}

\subsection*{Stiff Fluid $P=\rho$}

Now for ultra relativistic fluid we have $\alpha=1$ and the equation \eqref{WDeq} reduces to 
\begin{equation}\label{WD}
\left[\frac{\partial^{2}}{\partial\beta_{0}^{2}}-\frac{\partial^{2}}{\partial\beta_{+}^{2}}-\frac{\partial^{2}}{\partial\beta_{-}^{2}}\right]\psi = 24\imath\frac{\partial}{\partial T}\psi.
\end{equation}

Naively, \eqref{WD} seems to be non self-adjoint. But we can calculate deficiency index ($n_\pm$) of the operator, which turns out to be $0$, hence the operator is self-adjoint and the corresponding evolution is guaranteed to be unitary. The general solution to eq. \eqref{WD} has already been described in \cite{sridip}. In order to extract the physical informations, here we consider the solution corresponding to an eigenstate of both $p_{+}$ and $p_{-}$ with eigenvalues $k_{+}=k_{-}=0$, (we will see, what it does mean physically to have $k_{\pm}=0$ as we go along) which is given by
\begin{equation}
\label{26}
\psi =e^{\imath \left(\omega\beta_{0}-E T\right)}
\end{equation}
where energy, $E$ is given by $E=-\frac{\omega^{2}}{24}$. Now using linearity of the governing differential equation we obtain the following wave packet
\begin{equation}\label{26.1}
\Psi = \int e^{-\frac{\left(\omega +12\omega_{0}\right)^{2}}{4}} \psi d\omega,
\end{equation}
which upon integration and normalization, can be recast as
\begin{equation}
\label{27}
\Psi =\sqrt[4]{\frac{2}{\pi}}\frac{e^{-36\omega_{0}^{2}}}{\sqrt{1-\imath \frac{T}{6}}}\exp\left(-\frac{\left(\beta_{0}+\imath 6\omega_{0} \right)^{2}}{1-\imath \frac{T}{6}}\right),
\end{equation}
with a time dependent probability density function, given by
\begin{equation}
\label{28}
\Psi^{*}\Psi =\frac{\sqrt{2}}{\sqrt{\pi(1+\frac{T^{2}}{36})}}\exp\left(\frac{-2\left(\beta_{0}-\omega_{0}T\right)^{2}}{1+\frac{T^{2}}{36}}\right),
\end{equation}
which, on integration over the configuration space, yields a time independent value:
\begin{equation}\label{norm}
\int \Psi^{*}\Psi d\beta_{0}d\beta_{+}d\beta_{-}= 1\int d\beta_{-}d\beta_{+}= 1.\delta\left(0\right)\delta\left(0\right).
\end{equation}
 It deserves mention that presence of $\delta(0)$ indicates that the wavefunction belongs to the rigged Hilbert space just like the eigenfunction of momentum in elementary quantum mechanics.\\
 
The definition of norm used here \eqref{norm} is consistent with the definition of norm used for $\alpha\neq 1$ case, in the earlier work \cite{sridip}, in the sense, by properly taking limit $\alpha\rightarrow 1$ the norm used there reduce to the one, used above. For $\alpha\neq 1$, we have, for $\beta_{0}$ sector, \cite{sridip}
\begin{equation}\label{norm1}
||\psi ||_{\alpha\neq1}=\frac{3(1-\alpha)}{2}\int d\beta_{0} e^{\frac{3(1-\alpha)}{2}\beta_{0}}\psi^{*}\psi 
\end{equation}\label{norm2}
whereas for $\alpha=1$, the norm is defined as
\begin{equation}
|| \psi ||_{\alpha =1}=\int d\beta_0 \psi^{*}\psi = \lim_{\alpha\rightarrow 1}\frac{||\psi ||_{\alpha\neq 1}}{\frac{3}{2}(1-\alpha)}
\end{equation}

We note, the limit is well-defined since, \eqref{norm1} has a factor of $(1-\alpha)$. This also suggests that for $\alpha\neq 1$, we could have defined  norm without $\frac{3(1-\alpha)}{2}$ factor,  because physics is independent of such redefinition of a norm by a numerical factor. In that scenario, we don't even need to divide by the  $\frac{3(1-\alpha)}{2}$ factor in \eqref{norm2} to get the definition of norm for $\alpha=1$ case from $\alpha\neq 1$.\\

Having defined norm in proper way, now, for the wavefunction \eqref{27}, we can evaluate the expectation value of $\beta_{0}$, which behaves as follows:
\begin{equation}
\langle\beta_{0}\rangle =\omega_{0}T.
\end{equation}
while the expectation value of volume is given by:
\begin{equation}
\langle V\rangle =e^{\frac{9}{8}+3\omega_{0}T+\frac{T^{2}}{32}}.
\end{equation}

If we set $\omega_{0}=0$, the volume expansion can be depicted via figure 1. Although $\langle\beta_{0}\rangle$ is $0$ in this case (i.e with $\omega_{0}=0$), still we have expansion due to quantum fluctuation in the variable $\beta_{0}$. Another important aspect is that the volume assumes a finite value at T=0 and thereby avoids hitting the singularity as evident from figure 1. It deserves mention that by setting $\omega_{0}=0$ we do not loose any qualitative feature. As long as $\omega_{0}\geq0$, we do get an expansion. Only if $\omega_0$ becomes negative, we would get a contracting phase until $T=48|\omega_{0}|$, after which, the universe will start expanding. \\

 \begin{figure}[!ht]
 \centering
 \includegraphics[scale=0.5]{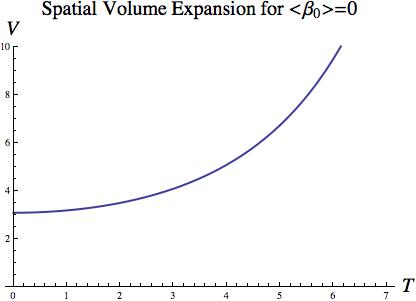}
 \caption{Behaviour of Volume with time for $\langle\beta_{0}\rangle=0$}
\end{figure}

Looking at the behavior of the wavefunction with time, as shown in figure 2, we observe that the peak of the Gaussian shifts toward right and spreads out for $T\geq 0$.
\begin{figure}[!ht]
\centering 
 \includegraphics[scale=0.5]{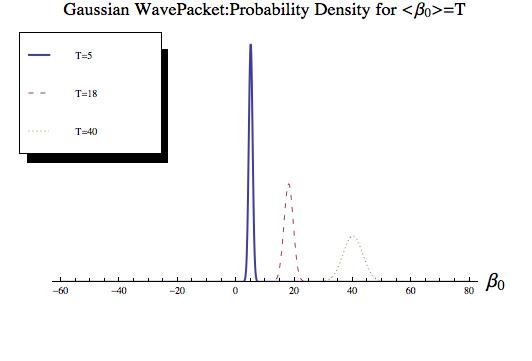}
 \caption{The Probability Distribution as a function of $\beta_{0}$ at different time $T$}
\end{figure}
 The spreading out happens because within the superposed solution there are different $\omega$ modes having different ``velocity", which ultimately leads to dispersion. The term ``velocity" corresponds to the rate of change of volume since the conjugate momentum to $\beta_{0}$ is actually proportional to the time derivative of volume factor. Thus, different $\omega$ modes may be assigned the physical interpretation of having different values for the rate of change of volume. This scenario can be compared to a Gaussian wavepacket of a free particle dispersing, in quantum mechanics. We note that the choice of Gaussian prefactor is based on the fact that it leads to a behaviour where the universe does not hit a singularity at T=0, we get expansion in the volume of the universe and we can nicely interpret the behaviour of the wavefunction. \\ 

The wavefunction, given by \eqref{27}, is  basically  Hartle-Hawking type, since we have superposition of both positive and negative $\omega$ modes. We can think of positive $\omega$ modes to describe an expanding universe whereas the negative modes can be thought of depicting a collapsing universe. Since these $\omega$ modes are weighted by Gaussian pre-factors, most of them gets suppressed and effectively we have an expanding universe.\\

As a side note, we observe $\Psi$ is also a solution to isotropic quantum FRW model with a spatially flat metric since $\Psi$ is independent of $\beta_{+},\beta_{-}$. Moreover, if we restrict to isotropic FRW model,  we do not need to integrate over $\beta_{+},\beta_{-}$, hence the delta functions would not appear in the norm \eqref{norm} and the wavefunction \eqref{27} perfectly belongs to the Hilbert space.\\

The fact that $\Psi$ does satisfy Hamiltonian constraint for quantum FRW universe, can be understood from inspecting $\beta_{0,\pm}$ in terms of scale factor. We note that
\begin{eqnarray}\label{PI}
\beta_{0}=\frac{1}{3}ln(abc)=\frac{1}{3}lnV,\\
\label{PI1}
\beta_{+}=\frac{1}{4}ln\left(\frac{ab}{c^{2}}\right),\\
\beta_{-}=\frac{1}{2\sqrt{3}}ln \left(\frac{a}{b}\right).
\end{eqnarray}
Hence, $\beta_{0}$ actually dictates the volume of the universe while the anisotropy is dictated by the variables $\beta_{\pm}$ and variables conjugate to them. If we have isotropic solution, the scale factors will have similar behaviour and we expect $k_{\pm}=0$ and $\langle\beta_{\pm}\rangle=0$.   Here we note that the $p_{\pm}$ sector admits plane wave like solution i.e $e^{\imath\left(k_{+}\beta_{+}+k_{-}\beta_{-}\right)}$ with $k_{\pm}=0$ and we do have $\langle\beta_{\pm}\rangle=0$ where the following limiting procedure is implicitly assumed 
\begin{equation}
\langle\beta_{\pm}\rangle =\frac{\int d\beta_{\pm}\beta_{\pm} e^{\imath k_{\pm}\beta_{\pm}}e^{-\imath k_{\pm}\beta_{\pm}}}{\int d\beta_{\pm} e^{\imath k_{\pm}\beta_{\pm}}e^{-\imath k_{\pm}\beta_{\pm}} }\equiv
\lim_{L\rightarrow\infty} \frac{\int_{-L}^{L} d\beta_{\pm}\beta_{\pm}}{\int_{-L}^{L} d\beta_{\pm}}
\end{equation}.
Hence, the solution is indeed isotropic in nature. The fact that the anisotropy is dictated by $\beta_{\pm}$ and $p_{\pm}$ is evident if we look at the scalar $\tilde{\sigma}^{2}=\frac{1}{2}\sigma^{\mu\nu}\sigma_{\mu\nu}$ found out of shear tensor $\sigma_{\mu\nu}$,
\begin{eqnarray}
\nonumber \tilde{\sigma}^{2} &=&
\frac{1}{12}\left[\left(\frac{\dot{g}_{11}}{g_{11}}-\frac{\dot{g}_{22}}{g_{22}}\right)^{2}
+ \left(\frac{\dot{g}_{22}}{g_{22}}-\frac{\dot{g}_{33}}{g_{33}}\right)^{2}+
\left(\frac{\dot{g}_{33}}{g_{33}}-\frac{\dot{g}_{11}}{g_{11}}\right)^{2}\right]\\ &=& \frac{1}{12e^{6\beta_{0}}}\left(\frac{p_{+}^{2}}{18}+\frac{p_{-}^{2}}{8}\right) ,
\end{eqnarray}\\

Now one way to make the model manifestly anisotropic is to take $\langle\beta_{\pm}\rangle\neq0$ or to make the momentum conjugate to $\beta_{\pm}$ nonzero. It deserves mention in this connection that the choice of $k_{\pm}=0$ may seem to be restricted, but in the next section, we will relax the condition and make $k_{+}\neq0$. To be precise, we will construct solution with nonzero $\langle\beta_{+}\rangle$ and $k_{+}$. In fact, it is quite trivial to extend the analysis to $k_{\pm}\neq 0$, we will get similar Gaussian solutions for $\beta_{\pm}$ sector as well. We have not done it since the essential physical features are conveyed with the restricted condition on $k_{\pm}$. This kind of choice has also been made earlier in the work of Alvarenga et. al. \cite{alvarenga3}.

\section{Explicit Anisotropy}
In the last section we constructed an isotropic solution. Hence, one can argue that the unitarity is obtained at the expense of anisotropy. In this section we present a solution, which is manifestly anisotropic. We consider eigenstate of $p_{-}$ with eigenvalue $k_{-}=0$ but let $\omega$ and $k_{+}$ to vary along with energy eigenvalue given by $E=-\frac{1}{24}\left(\omega^{2}-k_{+}^{2}\right)$. From the general solution $\psi=e^{\imath\left(\omega\beta_{0}+k_{+}\beta_{+}-ET\right)}$, we construct the following superposed solution :
\begin{equation}\label{01}
\Psi = \iint e^{-\frac{\left(\omega +12\omega_{0}\right)^{2}}{4}}e^{-\frac{\left(k_{+} -12k_{+0}\right)^{2}}{4}} \psi d\omega dk_{+},
\end{equation}
which, upon integration and normalization, yields
\begin{equation}\label{AS}
\Psi =\sqrt{\frac{2}{\pi}}\frac{e^{-36\left(\omega_{0}^{2}+k_{+0}^{2}\right)}}{\sqrt{1+\frac{T^{2}}{36}}}e^{-\frac{\left(\beta_{0}+\imath 6\omega_{0} \right)^{2}}{1-\imath \frac{T}{6}}}e^{-\frac{\left(\beta_{+}-\imath 6k_{+0} \right)^{2}}{1+\imath \frac{T}{6}}},
\end{equation}
with a time dependent probability density function, given by
\begin{equation}
\label{28}
\small{
\Psi^{*}\Psi =\frac{2}{\pi(1+\frac{T^{2}}{36})}\exp\left(\frac{-2\left[\left(\beta_{0}-\omega_{0}T\right)^{2}+\left(\beta_{+}-k_{+0}T\right)^{2}\right]}{1+\frac{T^{2}}{36}}\right)},
\end{equation}
which leads to a time independent norm, as evident from
\begin{equation}
\int d\beta_{0}d\beta_{+}d\beta_{-} \Psi^{*}\Psi = 1\delta(0).
\end{equation}
The $\delta(0)$ is there as we have a plane wave like solution for $k_{-}$ sector. Now, \eqref{AS} has the property
\begin{eqnarray}
\langle\beta_{0}\rangle & =&\omega_{0}T,\  \langle\beta_{+}\rangle = k_{+0}T\\
\langle \beta_{-}\rangle &=& 0
\end{eqnarray}

We have thus constructed an anisotropic solution explicitly since we have nonzero value for $\langle\beta_{+}\rangle$ and superposition of nonzero $k_{+}$ modes. Having $\langle\beta_{-}\rangle=0$ means that the scale factor $a$ and $b$ behaves similarly while non-zero expectation of $\beta_{+}$ means scale factor $c(t)$ behaves differently, as evident from equation \eqref{PI1}. The signature of anisotropy is also carried by expectation value of $\tilde{\sigma}^{2}$, which satisfies,
\begin{equation}
\langle \tilde{\sigma}^{2} \rangle = \frac{\langle e^{-6\beta_0}\rangle}{12}\left(\frac{\langle p_{+}^{2}\rangle}{18}+\frac{\langle p_{-}^{2}\rangle }{8}\right)=\frac{\langle e^{-6\beta_{0}}\rangle}{216}\neq 0
\end{equation} \\

While constructing the wavepacket \eqref{AS} , we have not put the constraint on the energy spectrum that it has to be negative on physical ground because we have Hamiltonian constraint \eqref{master} and we expect the energy of fluid to be positive definite\cite{sridip}. Now negative energy yields the constraint $\omega^{2}>k_{+}^{2}$. This, in turn, implies integration over a restricted interval in $(k_{+},\omega)$ space while obtaining the superposed solution, which would be difficult to inspect analytically. Nonetheless, the norm for this case remains finite. \\

\section{A more general fluid with $\alpha\neq 1$}
The solution to the \eqref{WDeq} for $\alpha\neq1$ has already been described in the reference \cite{sridip} and is given by:
\begin{equation}\label{solution}
\Psi =e^{\imath k_{+}\beta_{+}}e^{\imath k_{-}\beta_{-}}\phi_{a,b}e^{-\imath Et},
\end{equation} where $\phi_{a,b}$ are given in variable $\chi\equiv e^{\frac{3\left(1-\alpha\right)\beta_{0}}{2}}$ by 
\begin{eqnarray}\label{Master}
\phi_{a}(\chi)=\sqrt{\chi}\left[AH^{(2)}_{\imath g}(\lambda \chi)+BH_{\imath g}^{(1)}(\lambda \chi)\right],\\
\label{Master1}
\phi_{b}(\chi) =\sqrt{\chi}\left[AH^{(2)}_{\alpha}(\lambda \chi)+BH_{\alpha}^{(1)}(\lambda \chi)\right],
\end{eqnarray}
respectively for $\sigma > \frac{1}{4}$ and $\sigma < \frac{1}{4}$, where $g=-\imath\alpha$ $=\sqrt{\sigma-\frac{1}{4}}$ with $\sigma =\frac{4\left(k_{+}^{2}+k_{-}^{2}\right)}{9\left(1-\alpha \right)^{2}}$ so that $\phi_{a,b}$ satisfy,
\begin{equation}
\label{ne1}
-\frac{d^{2}\phi}{d\chi^{2}}-\frac{\sigma}{\chi^{2}}\phi =-E^{\prime}\phi .
\end{equation}
In both cases, the spectrum is given by $E^{\prime}=\frac{32}{3\left(1-\alpha \right)^{2}}E=-\lambda^{2}$. The self-adjoint extension guarantees that $\frac{B}{A}$ takes a value so as to conserve probability and makes the model unitary \cite{sridip}. For $\alpha\neq1$, it is technically difficult to find the expectation value of $\beta_0$ due to mathematical complexity of the integral involved
\begin{equation}
\langle\beta_{0}\rangle =\frac{2}{3\left(1-\alpha\right)}\frac{\int d\chi ln\left(\chi\right) |\phi |^{2}}{\int d\chi |\phi |^{2}},
\end{equation}
where $\phi$ is given by $\phi_{a}$ or $\phi_{b}$ for $\sigma >\frac{1}{4}$ or $\sigma <\frac{1}{4}$ respectively. Nonetheless, as we are dealing with energy eigenstate, this expectation value ought to be time independent trivially. \\

We now try to extract some physical information even without the integrated version. Asymptotically, the solutions \eqref{Master}, \eqref{Master1} become a linear combination of incoming and outgoing waves. Since, the eigenvalue of $p_{\chi}$ (momentum conjugate to the newly defined variable $\chi$) is proportional to rate of change of volume factor, $e^{3\beta_{0}}$, incoming wave can be thought of describing a collapsing universe with negative rate of change of volume whereas the outgoing wave describes an expanding universe. Hence, the solution \eqref{solution},being a superposition of expanding and collapsing universe, mimics Hartle-Hawking wavefunction \cite{AJP}. The probability conservation can now be physically thought of as the probability of collapsing being same as the probability of birth of a universe. \\

We can have Vilenkin type solution as well, all we need to do is to keep the outgoing solution while discarding the incoming wave solution so that the wavefunction becomes
\begin{eqnarray}
\phi_{a}(\chi)=\sqrt{\chi}\left[BH_{\imath g}^{(1)}(\lambda \chi)\right],\\
\phi_{b}(\chi) =\sqrt{\chi}\left[BH_{\alpha}^{(1)}(\lambda \chi)\right].
\end{eqnarray}
Previously, the boundary condition has imposed restriction on the ratio of A and B. Now since, we have $A=0$, the boundary condition enforces constraint on the energy spectra. It can be shown for $\sigma<1/4$, the condition coming from self-adjoint extension \cite{griffiths, sridip}, given by
\begin{equation}\label{ec}
\alpha\frac{\pi}{2}+\imath\alpha ln\left(\frac{\lambda x_{0}}{2}\right) +arg[\Gamma\left(1+\alpha\right)]=n\pi,
\end{equation}
(where $x_{0}$ characterizes the self-adjoint extension, elucidated in \cite{griffiths}) is satisfied if 
\begin{eqnarray}
\label{C1}
x_{0}&=&|x_{0}|e^{\imath \frac{\pi}{2} \theta},\\
\label{C2}
\left(\frac{\lambda |x_{0}|}{2}\right)&=&1,\\
\label{C3}
\alpha\left(1-\theta\right) &=& 2n,\\
\label{C4}
\sigma \equiv \frac{4\left(k_{+}^{2}+k_{-}^{2}\right)}{9\left(1-\alpha \right)^{2}} & =&\frac{1}{4}- \frac{4n^{2}}{\left(1-\theta \right)^{2}}.
\end{eqnarray} 
The condition \eqref{C2} implies existence of unique energy eigenstate with energy eigenvalue, given by
\begin{equation}
E^{\prime}=-\frac{4}{|x_{0}|^{2}}.
\end{equation}

Similar analysis reveals that for $\sigma\geq \frac{1}{4}$, $\theta$ must be equal to 1 i.e $x_{0}=\imath|x_{0}|$, which yields a unique energy eigenstate with $E^{\prime}=-\frac{4}{\gamma^{2}|x_{0}|^{2}}$ for $\sigma=\frac{1}{4}$ (where $ln(\gamma)$ is Euler's constant) while for $\sigma>\frac{1}{4}$, we have 
a bunch of energy eigenvalues $E^{\prime}=-\lambda^{2}$ satisfying
\begin{equation}
gln\left(\frac{\lambda |x_{0}|}{2}\right) - arg [\Gamma\left(1+\imath g\right)]=n\pi.
\end{equation} 

\subsection*{Unphysical positive energy states}
In \cite{sridip}, it has been argued that observed universe with positive energy implies that the fluid sector has negative energy (because of the Hamiltonian constraint \eqref{master}), which is not physical. Here we will argue from the point of view of gravity why the positive energy eigenvalue does not make sense physically.

We have the following solution with positive eigenvalue for $\sigma>\frac{1}{4}$ :
\begin{equation}
\label{59}
\phi = \lambda \sqrt{\frac{2\sinh\left(\pi g\right)}{\pi g}}\sqrt{\chi} K_{ig}\left(\lambda \chi\right).
\end{equation}
where $g=\sqrt{\sigma-\frac{1}{4}}$ (the wavefunction is shown in figure 3 below) with energy eigenvalue given by $E^{\prime}=\lambda^{2}$.

  \begin{figure}[!h]
  \centering
 \includegraphics[scale=0.5]{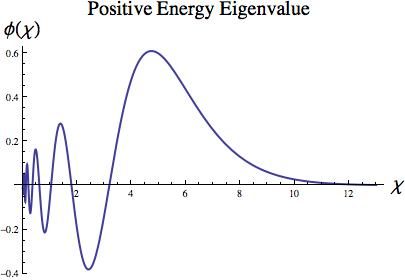}
 \caption{Positive Energy Eigenstate with $E^{\prime}=1,g=6$}
\end{figure}

The solution \eqref{59} gets exponentially damped for high $\chi$ (i.e high $\beta_{0}$) value. It implies that we have exponentially damped probability for having larger $\beta_{0}$ i.e larger volume which is not physical for an observed expanding universe.The same conclusion can be drawn even if we do not have the explicit form of the wavefunction but we look at governing differential equation \eqref{ne1} in asymptotic limit and realise that for arbitrary large $\chi$ , it becomes
\begin{equation}
-\frac{d^{2}\phi}{d\chi^{2}}= -E^{\prime}\phi ,
\end{equation} 
which does imply exponential decay in $\chi$.\\

This comes as a sanity check for our theory. Positive eigenvalue for the gravity sector implies negative fluid energy which indicates a `trapped' universe, where the volume has a very low probability to get larger than some specified value which depends on the energy. It deserves mention that $\sigma<\frac{1}{4}$ does not admit any positive energy eigenvalue \cite{griffiths} and this fact hints at we have two different regime of solution around the critical point $\sigma=\frac{1}{4}$ \cite{Camblong, Nisoli}.

\section{Conclusion}
The work reinforces the idea presented in \cite{sridip,sridip2} that the alleged non unitarity in a quantum anisotropic cosmological model has nothing to do with hyperbolic nature of kinetic term, rather it is intimately linked to co-ordinate choice, which in turn affects the operator ordering.  Although we do not know the physical reason behind why we do have to choose this ordering, yet, with proper ordering, we restore the unitarity and thereby make the model worthy of being investigated critically. This, in turn, enables us to have physical insight and predictions from this model, which is cured from alleged non unitarity.\\

It has been shown in this work that the universe does not hit any singularity at $T=0$. The work sheds light on the nature of variables $\beta_{\pm}$ and $p_{\pm}$, which encode the anisotropic nature of the solution and thereby exhibit that the solution obtained is indeed anisotropic in nature. Furthermore, we physically interpret the wavefunction in terms of collapsing and expanding universe and thereby draw a possible parallelism between our wavefunction with Hartle-Hawking type $\&$ Vilenkin type wavefunctions. \\

The solution for $\alpha\neq1$, given in reference \cite{sridip}, is shown to be a Hartle-Hawking type wavefunction. In this work, we extend our analysis and construct a Vilenkin type wavefunction for which the self-adjoint extension puts constraint on energy eigenspectra. For $\sigma>\frac{1}{4}$, we have a bunch of eigenvalues while  for $\sigma \leq\frac{1}{4}$ case, we have a unique solution. The study of this transition with possible breaking of conformal symmetry when the system passes through the critical point $\sigma=\frac{1}{4}$ is beyond the scope of this work yet it deserves mention that similar kind of phase transition associated with inverse square potential has already been known\cite{Camblong} and mimics the BKT phase transition \cite{Nisoli}. Moreover, the work gives another perspective to the physical impossibility of having positive energy eigenstate of Hamiltonian of gravity sector in terms of cosmic evolution.\\

We also conclusively show that the anisotropy is dictated by the variables $\beta_{\pm}$ whereas the ordering ambiguity is in the variable $\beta_{0}$, which dictates the evolution of volume. Hence, it lends support to the idea presented in \cite{sridip,sridip2} that the pathology of non unitarity is not a generic feature of anisotropic models, it may crop up even in isotropic case \cite{sridip2}, the problem rather lies in choosing the appropriate co-ordinate and operator ordering for the variable which describes the evolution of volume.\\
 
\section*{Acknowledgement}
The author would like to acknowledge a debt of gratitude to Professor Narayan Banerjee without whose active support, the work could not have been possible. The author would also like to thank Diptarka Das and Shouvik Ganguly for a stimulating discussion. \\

\end{document}